\documentclass[useAMS,usenatbib]{mn2e}
\usepackage{tabularx}
\usepackage{xspace}
\usepackage{graphicx}
\newcommand{\ignore}[1]{}
\providecommand{\ao}{}

\renewcommand{\ao}{adaptive optics (AO)\renewcommand{\ao}{AO\xspace}\renewcommand{\Ao}{AO\xspace}\xspace}
\newcommand{\Ao}{Adaptive optics (AO)\renewcommand{\ao}{AO\xspace}\renewcommand{\Ao}{AO\xspace}\xspace}
\newcommand{\wfs}{wavefront sensor (WFS)\renewcommand{\wfs}{WFS\xspace}\renewcommand{\wfss}{WFSs\xspace}\xspace}
\newcommand{\wfss}{wavefront sensors (WFSs)\renewcommand{\wfs}{WFS\xspace}\renewcommand{\wfss}{WFSs\xspace}\xspace}
\newcommand{\shwfs}{Shack-Hartmann \wfs (SHWFS)\renewcommand{\shwfs}{SHWFS\xspace}\xspace}
\newcommand{\dm}{deformable mirror (DM)\renewcommand{\dm}{DM\xspace}\renewcommand{\dms}{DMs\xspace}\renewcommand{\Dms}{DMs\xspace}\renewcommand{\Dm}{DM\xspace}\xspace}
\newcommand{\dms}{deformable mirrors (DMs)\renewcommand{\dm}{DM\xspace}\renewcommand{\dms}{DMs\xspace}\renewcommand{\Dms}{DMs\xspace}\renewcommand{\Dm}{DM\xspace}\xspace}
\newcommand{\Dms}{Deformable mirrors (DMs)\renewcommand{\dm}{DM\xspace}\renewcommand{\dms}{DMs\xspace}\renewcommand{\Dms}{DMs\xspace}\renewcommand{\Dm}{DM\xspace}\xspace}
\newcommand{\Dm}{Deformable mirror (DM)\renewcommand{\dm}{DM\xspace}\renewcommand{\dms}{DMs\xspace}\renewcommand{\Dms}{DMs\xspace}\renewcommand{\Dm}{DM\xspace}\xspace}

\newcommand{\lqg}{linear-quadratic-gaussian (LQG)\renewcommand{\lqg}{LQG\xspace}\xspace}
\newcommand{\shs}{Shack-Hartmann sensor (SHS)\renewcommand{\shs}{SHS\xspace}\renewcommand{\shss}{SHSs\xspace}\xspace}
\newcommand{\shss}{Shack-Hartmann sensors (SHSs)\renewcommand{\shs}{SHS\xspace}\renewcommand{\shss}{SHSs\xspace}\xspace}
\newcommand{\lgs}{laser guide star (LGS)\renewcommand{\lgs}{LGS\xspace}\renewcommand{\Lgs}{LGS\xspace}\renewcommand{\lgss}{LGSs\xspace}\xspace}
\newcommand{\lgss}{laser guide stars (LGSs)\renewcommand{\lgs}{LGS\xspace}\renewcommand{\Lgs}{LGS\xspace}\renewcommand{\lgss}{LGSs\xspace}\xspace}
\newcommand{\Lgs}{Laser guide star (LGS)\renewcommand{\lgs}{LGS\xspace}\renewcommand{\Lgs}{LGS\xspace}\renewcommand{\lgss}{LGSs\xspace}\xspace}

\newcommand{\Ngs}{Natural guide star (NGS)\renewcommand{\ngs}{NGS\xspace}\renewcommand{\Ngs}{NGS\xspace}\renewcommand{\ngss}{NGSs\xspace}\xspace}
\newcommand{\ngs}{natural guide star (NGS)\renewcommand{\ngs}{NGS\xspace}\renewcommand{\Ngs}{NGS\xspace}\renewcommand{\ngss}{NGSs\xspace}\xspace}
\newcommand{\ngss}{natural guide stars (NGSs)\renewcommand{\ngs}{NGS\xspace}\renewcommand{\Ngs}{NGS\xspace}\renewcommand{\ngss}{NGSs\xspace}\xspace}
\newcommand{\mems}{Micro-Electro-Mechanical Systems (MEMS)\renewcommand{\mems}{MEMS\xspace}\xspace}
\newcommand{\snr}{signal to noise ratio (SNR)\renewcommand{\snr}{SNR\xspace}\xspace}
\newcommand{\Moao}{Multi-object \ao (MOAO)\renewcommand{\moao}{MOAO\xspace}\renewcommand{\Moao}{MOAO\xspace}\xspace}
\newcommand{\moao}{multi-object \ao (MOAO)\renewcommand{\moao}{MOAO\xspace}\renewcommand{\Moao}{MOAO\xspace}\xspace}
\newcommand{\mcao}{multi-conjugate adaptive optics (MCAO)\renewcommand{\mcao}{MCAO\xspace}\xspace}
\newcommand{\ltao}{laser tomographic \ao (LTAO)\renewcommand{\ltao}{LTAO\xspace}\xspace}
\newcommand{\cpu}{central processing unit (CPU)\renewcommand{\cpu}{CPU\xspace}\renewcommand{\cpus}{CPUs\xspace}\xspace}
\newcommand{\cpus}{central processing units (CPUs)\renewcommand{\cpu}{CPU\xspace}\renewcommand{\cpus}{CPUs\xspace}\xspace}
\newcommand{\psf}{point spread function (PSF)\renewcommand{\psf}{PSF\xspace}\renewcommand{\psfs}{PSFs\xspace}\renewcommand{\Psf}{PSF\xspace}\xspace}
\newcommand{\psfs}{point spread functions (PSFs)\renewcommand{\psf}{PSF\xspace}\renewcommand{\psfs}{PSFs\xspace}\renewcommand{\Psf}{PSF\xspace}\xspace}
\newcommand{\Psf}{Point spread function (PSF)\renewcommand{\psf}{PSF\xspace}\renewcommand{\psfs}{PSFs\xspace}\renewcommand{\Psf}{PSF\xspace}\xspace}
\newcommand{\fpga}{field programmable gate array (FPGA)\renewcommand{\fpga}{FPGA\xspace}\renewcommand{\fpgas}{FPGAs\xspace}\xspace}
\newcommand{\fpgas}{field programmable gate arrays (FPGAs)\renewcommand{\fpga}{FPGA\xspace}\renewcommand{\fpgas}{FPGAs\xspace}\xspace}
\newcommand{\sor}{successive over-relaxation (SOR)\renewcommand{\sor}{SOR\xspace}\xspace}
\newcommand{\fdpcg}{Fourier domain pre-conditioned gradient (FDPCG)\renewcommand{\fdpcg}{FDPCG\xspace}\xspace}
\newcommand{\map}{maximum a-posteriori (MAP)\renewcommand{\map}{MAP\xspace}\xspace}

\newcommand{\elt}{Extremely Large Telescope (ELT)\renewcommand{\elt}{ELT\xspace}\renewcommand{\elts}{ELTs\xspace}\renewcommand{\eelt}{European ELT (E-ELT)\renewcommand{\eelt}{E-ELT\xspace}\xspace}\xspace}

\newcommand{\elts}{Extremely Large Telescopes (ELTs)\renewcommand{\elt}{ELT\xspace}\renewcommand{\elts}{ELTs\xspace}\renewcommand{\eelt}{European ELT (E-ELT)\renewcommand{\eelt}{E-ELT\xspace}\xspace}\xspace}

\newcommand{\eelt}{European Extremely Large Telescope (E-ELT)\renewcommand{\eelt}{E-ELT\xspace}\renewcommand{\elt}{ELT\xspace}\renewcommand{\elts}{ELTs\xspace}\xspace}

\newcommand{\dugall}{Durham University generalised adaptive optics laser laboratory (DUGALL)\renewcommand{\dugall}{DUGALL\xspace}\xspace}
\newcommand{\fwhm}{full-width at half-maximum (FWHM)\renewcommand{\fwhm}{FWHM\xspace}\xspace}
\newcommand{\wht}{William Herschel Telescope (WHT)\renewcommand{\wht}{WHT\xspace}\xspace}
\newcommand{\emccd}{electron multiplying CCD (EMCCD)\renewcommand{\emccd}{EMCCD\xspace}\renewcommand{\emccds}{EMCCDs\xspace}\xspace}
\newcommand{\emccds}{electron multiplying CCDs (EMCCDs)\renewcommand{\emccd}{EMCCD\xspace}\renewcommand{\emccds}{EMCCDs\xspace}\xspace}
\newcommand{\dasp}{Durham \ao simulation platform (DASP)\renewcommand{\dasp}{DASP\xspace}\renewcommand{\thedasp}{DASP\xspace}\renewcommand{\Thedasp}{DASP\xspace}\xspace}
\newcommand{\thedasp}{the Durham \ao simulation platform (DASP)\renewcommand{\dasp}{DASP\xspace}\renewcommand{\thedasp}{DASP\xspace}\renewcommand{\Thedasp}{DASP\xspace}\xspace}
\newcommand{\Thedasp}{The Durham \ao simulation platform (DASP)\renewcommand{\dasp}{DASP\xspace}\renewcommand{\thedasp}{DASP\xspace}\renewcommand{\Thedasp}{DASP\xspace}\xspace}
\newcommand{\mpi}{Message Passing Interface (MPI)\renewcommand{\mpi}{MPI\xspace}\xspace}
\newcommand{\smp}{symmetric multi-processing (SMP)\renewcommand{\smp}{SMP\xspace}\xspace}
\newcommand{\svd}{singular value decomposition (SVD)\renewcommand{\svd}{SVD\xspace}\xspace}
\newcommand{\gpu}{graphics processing unit (GPU)\renewcommand{\gpu}{GPU\xspace}\renewcommand{\gpus}{GPUs\xspace}\xspace}
\newcommand{\gpus}{graphics processing units (GPUs)\renewcommand{\gpu}{GPU\xspace}\renewcommand{\gpus}{GPUs\xspace}\xspace}
\newcommand{\fft}{fast Fourier transform (FFT)\renewcommand{\fft}{FFT\xspace}\xspace}
\newcommand{\ifu}{integral field unit (IFU)\renewcommand{\ifu}{IFU\xspace}\xspace}
\newcommand{\darc}{the Durham \ao real-time controller (DARC)\renewcommand{\darc}{DARC\xspace}\renewcommand{\Darc}{DARC\xspace}\xspace}
\newcommand{\Darc}{The Durham \ao real-time controller (DARC)\renewcommand{\darc}{DARC\xspace}\renewcommand{\Darc}{DARC\xspace}\xspace}
\newcommand{\cots}{commercial off-the-shelf (COTS)\renewcommand{\cots}{COTS\xspace}\xspace}
\newcommand{\rtcp}{real-time control pipeline (RTCP)\renewcommand{\rtcp}{RTCP\xspace}\xspace}
\newcommand{\rms}{root-mean-square (RMS)\renewcommand{\rms}{RMS\xspace}\xspace}
\newcommand{\sFPDP}{serial Front Panel Data Port (sFPDP)\renewcommand{\sFPDP}{sFPDP\xspace}\xspace}
\newcommand{\wpu}{wavefront processing unit (WPU)\renewcommand{\wpu}{WPU\xspace}\xspace}

\newcommand{\rtcs}{real-time control system (RTCS)\renewcommand{\rtcs}{RTCS\xspace}\renewcommand{\rtcss}{RTCSs\xspace}\xspace}
\newcommand{\rtcss}{real-time control systems (RTCSs)\renewcommand{\rtcs}{RTCS\xspace}\renewcommand{\rtcss}{RTCSs\xspace}\xspace}
\newcommand{\eso}{European Southern Observatory (ESO)\renewcommand{\eso}{ESO\xspace}\renewcommand{\theeso}{ESO\xspace}\xspace}
\newcommand{\theeso}{\renewcommand{\theeso}{ESO\xspace}the \eso}
\newcommand{\scao}{single conjugate \ao (SCAO)\renewcommand{\scao}{SCAO\xspace}\renewcommand{\Scao}{SCAO\xspace}\xspace}
\newcommand{\Scao}{Single conjugate \ao (SCAO)\renewcommand{\scao}{SCAO\xspace}\renewcommand{\Scao}{SCAO\xspace}\xspace}
\newcommand{\glao}{ground layer \ao (GLAO)\renewcommand{\glao}{GLAO\xspace}\xspace}
\newcommand{\eagle}{ELT Adaptive optics for GaLaxy Evolution (EAGLE)\renewcommand{\eagle}{EAGLE\xspace}\xspace}
\newcommand{\maory}{multi-conjugate \ao relay for the \eelt (MAORY)\renewcommand{\maory}{MAORY\xspace}\xspace}
\newcommand{\muse}{Multi Unit Spectroscopic Explorer (MUSE)\renewcommand{\muse}{MUSE\xspace}\xspace}
\newcommand{\vlt}{Very Large Telescope (VLT)\renewcommand{\vlt}{VLT\xspace}\xspace}
\newcommand{\ccd}{CCD\xspace}
\newcommand{\tmt}{Thirty Metre Telescope (TMT)\renewcommand{\tmt}{TMT\xspace}\xspace}
\newcommand{\xao}{eXtreme \ao (XAO)\renewcommand{\xao}{XAO\xspace}\xspace}

\newcommand{\vla}{Very Large Array (VLA)\renewcommand{\vla}{VLA\xspace}\xspace}
\newcommand{\jwst}{{\em James Webb Space Telescope} \citep[JWST,][]{jwst}\renewcommand{\jwst}{{\em JWST}\xspace}\xspace}
\newcommand{\hst}{{\em Hubble Space Telescope (HST)}\renewcommand{\hst}{{\em HST}\xspace}\xspace}
\newcommand{\ifss}{integral-field spectrographs (IFSs)\renewcommand{\ifss}{IFSs\xspace}\renewcommand{\ifs}{IFS\xspace}\xspace}
\newcommand{\ifs}{integral-field spectrograph (IFS)\renewcommand{\ifss}{IFSs\xspace}\renewcommand{\ifs}{IFS\xspace}\xspace}
\newcommand{\ifus}{integral field units (IFUs)\renewcommand{\ifus}{IFUs\xspace}\xspace}
\newcommand{\mos}{multi-object spectrograph (MOS)\renewcommand{\mos}{MOS\xspace}\xspace}
\newcommand{\goodss}{Great Observatories Origins Deep Survey (GOODS)-S\renewcommand{\goodss}{GOODS-S\xspace}\xspace}
\newcommand{\goods}{Great Observatories Origins Deep Survey (GOODS)\renewcommand{\goods}{GOODS\xspace}\xspace}
\newcommand{\scmos}{scientific CMOS (sCMOS)\renewcommand{\scmos}{sCMOS\xspace}\xspace}
\newcommand{\aof}{Adaptive Optics Facility (AOF)\renewcommand{\aof}{AOF\xspace}\xspace}
\newcommand{\dsp}{digital signal processor (DSP)\renewcommand{\dsp}{DSP\xspace}\renewcommand{\dsps}{DSPs\xspace}\xspace}
\newcommand{\dsps}{digital signal processors (DSPs)\renewcommand{\dsp}{DSP\xspace}\renewcommand{\dsps}{DSPs\xspace}\xspace}
\newcommand{\capi}{Coherent Accelerator Processor Interface (CAPI)\renewcommand{\capi}{CAPI\xspace}\xspace}
\newcommand{\qe}{quantum efficiency (QE)\renewcommand{\qe}{QE\xspace}\xspace}
\newcommand{\numa}{non-uniform memory access (NUMA)\renewcommand{\numa}{NUMA\xspace}\xspace}
\newcommand{\uav}{unmanned aerial vehicle (UAV)\renewcommand{\uav}{UAV\xspace}\renewcommand{\uavs}{UAVs\xspace}\xspace}
\newcommand{\uavs}{unmanned aerial vehicles (UAVs)\renewcommand{\uav}{UAV\xspace}\renewcommand{\uavs}{UAVs\xspace}\xspace}

\newcommand{\ncpa}{non-common path aberration (NCPA)\renewcommand{\ncpa}{NCPA\xspace}\renewcommand{\ncpas}{NCPAs\xspace}\xspace}
\newcommand{\ncpas}{non-common path aberrations (NCPA)\renewcommand{\ncpa}{NCPA\xspace}\renewcommand{\ncpas}{NCPAs\xspace}\xspace}
\newcommand{\sdk}{software developers kit (SDK)\renewcommand{\sdk}{SDK\xspace}\renewcommand{\sdks}{SDKs\xspace}\xspace}
\newcommand{\sdks}{software developers kits (SDKs)\renewcommand{\sdk}{SDK\xspace}\renewcommand{\sdks}{SDKs\xspace}\xspace}
\newcommand{\dac}{digital to analogue converter (DAC)\renewcommand{\dac}{DAC\xspace}\xspace}
\newcommand{\nda}{non-disclosure agreement (NDA)\renewcommand{\nda}{NDA\xspace}\xspace}
\newcommand{\polc}{pseudo-open-loop control (POLC)\renewcommand{\polc}{POLC\xspace}\xspace}
\newcommand{\udp}{User Datagram Protocol (UDP)\renewcommand{\udp}{UDP\xspace}\xspace}

\newcommand{\usb}{universal serial bus (USB)\renewcommand{\usb}{USB\xspace}\xspace}
\usepackage{amssymb}

\title[Experience with WFS and DM interfaces for wide-field AO]{Experience with wavefront sensor and deformable mirror interfaces for wide-field
  adaptive optics systems}

\author[A.\ G.\ Basden et al.]{A.\ G.\ Basden$^{1}$\thanks{E-mail:
    a.g.basden@durham.ac.uk (AGB)},
D.\ Atkinson,$^{2}$ 
N.\ A.\ Bharmal,$^{1}$
U.\ Bitenc,$^{1}$
M.\ Brangier,$^{3}$ 
T.\ Buey,$^{3}$
\newauthor
T.\ Butterley,$^{1}$ 
D.\ Cano,$^{11}$ 
F.\ Chemla,$^{12}$ 
P.\ Clark,$^{1}$ 
M.\ Cohen,$^{12}$ 
J.-M.\ Conan,$^{4}$ 
\newauthor
F.\ J.\ de Cos,$^{5}$ 
C.\ Dickson,$^{2}$ 
N.\ A.\ Dipper,$^{1}$ 
C.\ N.\ Dunlop,$^{1}$ 
P.\ Feautrier,$^{13}$
T.\ Fusco,$^{4, 6}$ 
\newauthor
J.\ L.\ Gach,$^{6}$
E.\ Gendron,$^{3}$ 
D.\ Geng,$^{1}$ 
S.\ J.\ Goodsell,$^{1}$ 
D.\ Gratadour,$^{3}$ 
\newauthor
A.\ H.\ Greenaway,$^{7}$ 
A.\ Guesalaga,$^{8}$ 
C.\ D.\ Guzman,$^{8}$ 
D.\ Henry,$^{2}$ 
D.\ Holck,$^{1}$ 
Z.\ Hubert,$^{3}$ 
\newauthor
J.\ M.\ Huet,$^{12}$ 
A.\ Kellerer,$^{9}$ 
C.\ Kulcsar,$^{10}$ 
P.\ Laporte,$^{12}$ 
B.\ Le Roux,$^{6}$ 
N.\ Looker,$^{1}$ 
\newauthor
A.\ J.\ Longmore,$^{2}$ 
M.\ Marteaud,$^{3}$ 
O.\ Martin,$^{3}$ 
S.\ Meimon,$^{4}$  
C.\ Morel,$^{3}$
T.\ J.\ Morris,$^{1}$ 
\newauthor
R.\ M.\ Myers,$^{1}$ 
J.\ Osborn,$^{1}$ 
D.\ Perret,$^{3}$ 
C.\ Petit,$^{4}$ 
H.\ Raynaud,$^{10}$ 
\newauthor
A.\ P.\ Reeves,$^{1}$ 
G.\ Rousset,$^{3}$ 
F.\ Sanchez Lasheras,$^{5}$ 
M.\ Sanchez Rodriguez,$^{5}$ 
J.\ D.\ Santos,$^{5}$ 
\newauthor
A.\ Sevin,$^{3}$ 
G.\ Sivo,$^{10}$ 
E.\ Stadler,$^{13}$
B.\ Stobie,$^{2}$ 
G.\ Talbot,$^{1}$ 
S.\ Todd,$^{2}$ 
F.\ Vidal,$^{3}$ 
\newauthor
E.\ J.\ Younger,$^{1}$ 
\\
$^{1}$Department of Physics, South Road, Durham, DH1 3LE,
UK,\\ $^{2}$UKATC, Blackford Hill, Edinburgh, UK,\\ $^{3}$LESIA, Observatoire de
Paris, CNRS, Univ. Paris Diderot, France,\\ $^{4}$ONERA, 29 Avenue de la Division Leclerc, 92320
Chatillon, Paris, France,\\
$^{5}$Oviedo University, Calle San Francisco, 1, 33003, Oviedo, Spain,\\ $^{6}$Laboratoire d'Astrophysique de
Marseille, Rue Frederic Joliot Curie, 13013, Marseille,
France,\\ $^{7}$Herriot Watt University, Edinburgh Campus, Edinburgh,
EH14 4AS, UK,\\
$^{8}$Pontificia Universidad Católica de Chile, Avda Libertador Bernardo O'Higgins 340, Santiago , Chile,\\ $^{9}$Cambridge
University, Trinity Lane, Cambridge, CB2 1TN, UK,\\
$^{10}$Institut d'Optique Graduate School, 2 Avenue Augustin Fresnel,
91127 Palaiseau, Paris, France,\\ $^{11}$Issac Newton Group, Edificio Mayantigo, Calle
Alvarez Abreu, 70, E-38700 Santa Cruz de la Palma, Canary Islands,
Spain,\\ $^{12}$GEPI, Obs. de Paris, CNRS, Univ. Paris Diderot,
France.\\
$^{13}$IPAG, Institut de Planetologie et d'Astrophysique de Grenoble,
BP 53 F-38041, Grenoble, Cedex 9, France.
}

\newcommand{\sparta}{SPARTA\xspace}
\newcommand{\sfpdp}{serial Front Panel Data Port (sFPDP)\renewcommand{\sfpdp}{sFPDP\xspace}\xspace}
\newcommand{\pfpdp}{parallel Front Panel Data Port (pFPDP)\renewcommand{\pfpdp}{pFPDP\xspace}\xspace}
\newcommand{\dmc}{deformable mirror controller (DMC)\renewcommand{\dmc}{DMC\xspace}\xspace}
\newcommand{\INT}{Isaac Newton Telescope (INT)\renewcommand{\INT}{INT\xspace}\xspace}

\begin{document}
\maketitle

\begin{abstract}
Recent advances in adaptive optics (AO) have led to the implementation of
wide field-of-view AO systems.  A number of wide-field AO systems are
also planned for the forthcoming Extremely Large Telescopes.  Such
systems have multiple wavefront sensors of different types, and
usually multiple deformable mirrors (DMs).

Here, we report on our experience integrating cameras and DMs with the
real-time control systems of two wide-field AO systems.  These are
CANARY, which has been operating on-sky since 2010, and DRAGON, which
is a laboratory adaptive optics real-time demonstrator instrument.  We
detail the issues and difficulties that arose, along with the
solutions we developed.  We also provide recommendations for
consideration when developing future wide-field AO systems.
\end{abstract}

\ignore{
B1 was an Andor

LGS commissioning was a Pockels cell

We intended to use the Scimeasure all along, but it wasn't working during
commissioning or B1 so we went back to the Pockels. We could only get a single
LGS through the Pockels with any kind of contrast ratio so we had to do an
intermediate B1 phase.

B2 we started with the Scimeasure, but then it blew up between B2 and C1.

C1. Bobcat first, OCAM with shit gain and the eye of sauron

C2. All happy. OCAM with improved triggering, shuttering, bias and an increased
FOV on the lenslet to concentrate the light on fewer pixels.
}

\section{Introduction}
The forthcoming \elts \citep{eelt,tmt,gmt} all rely on \ao systems
\citep{adaptiveoptics} to provide atmospheric turbulence compensation
and compensation for telescope vibrations due to wind loading and
other motion.  These \ao systems are essential to
allow scientific goals requiring high resolution imaging and
spectroscopy to be met.  The vast majority of astronomical
observations made with these telescopes will use wide-field-of-view
\ao systems, including \mcao, \glao, \moao and \ltao.  These systems all use
information from multiple \wfss to provide a tomographic
reconstruction of the Earth's atmospheric turbulence.  

All current wide-field \ao systems on existing telescopes have seen
first light within the past decade, with facility class instruments
such as GeMS \citep{2012SPIE.8447E..0IRshort} only undergoing commissioning in the past
two years.  Therefore current operational experience is limited, and
each system comes with its own complexities and problems.

The CANARY instrument \citep{canaryshort,canaryresultsshort} on the \wht is the most
advanced wide-field \ao test-bed worldwide with an on-sky capability.  It has been
operated in \moao, \glao and \ltao modes, in addition to \scao for
comparative purposes.  CANARY relies on both natural and laser guide
stars, and has operated with both low and high resolution wavefront
sensing modes.  CANARY has been under continuous development having
seen four major phases of operation, and with a Sodium \lgs scheduled
for commissioning in mid-2016.  One outcome of this continuous
development is that we have amassed extensive experience interfacing
different \wfs cameras with the CANARY real-time control
system, and have developed techniques to handle \wfs synchronisation
in the presence of partial, corrupted or missing \wfs frames, different \wfs
interfaces, and unreliable camera
interfaces.  CANARY has also operated with several different
\dms.  In this paper, we discuss our experiences with \wfs and \dm
interfaces to the CANARY real-time control system.

The DRAGON \ao test-bench at Durham University \citep{dragon} is a
real-time wide-field \ao demonstrator, which is used to explore
wide-field \ao techniques with high order \wfss at 4--8~m telescope
scales.  This system models multiple \ngss and \lgss (including spot
elongation and laser launch up-link through turbulence) and uses  woofer-tweeter \dm control
\citep{woofertweeter}.  Although the configuration of DRAGON is more
permanent than that of CANARY (i.e.\ we will not be adding additional
\wfss or \dms in the foreseeable future), it still provides us with
experience with interfacing \wfs cameras and \dms to the real-time
control system, and we discuss this experience here.  

Both DRAGON and CANARY share a common real-time control system,
\darc, which is \cpu-based with optional hardware acceleration
facilities, including \gpus and \fpgas \citep{basden9}.  \darc is well
suited for \elt-scale operation \citep{basden11}.  This common base allows
\wfs and \dm interfaces to be shared between systems where necessary
with little additional effort, and a \cpu-based software on commodity hardware
greatly simplifies the addition of new \wfs camera and \dm interfaces.

In \S2 we provide a historical overview of the different CANARY
operational phases, and of the \wfs and \dm interfaces developed.
This gives a historical narrative of how the world-leading CANARY
instrument was developed and built.  We also provide this information
for DRAGON.  In \S3, we discuss the challenges we experienced, and the
techniques used to overcome these.  We conclude in \S4.

\section{Interfaces developed for CANARY and DRAGON}
CANARY was first operated on-sky in 2010, with 4 \ngss, a single \dm
and a tip-tilt mirror, performing \scao, \moao and \glao correction
(phase A).  During 2011, Rayleigh \lgs commissioning was carried out,
though without any \ao correction (phase B0).  In 2012, CANARY was
operated with a single Rayleigh \lgs and 4 \ngs (phase B1), and during 2013,
was upgraded to include 4 \lgs (phase B2).  In 2014, CANARY was
reconfigured to operate in \ltao mode (phase C1), and in 2015, 
a second 241-actuator \dm was added to create a split open/closed loop
system including woofer-tweeter operation (phase C2).  Both the \lgs
and \scao \ngs \wfss were upgraded from $7\times7$ to $14\times14$
sub-apertures at this point.  In 2016, a Sodium \lgs (launched far
from the telescope axis, up to 40~m away) will replace the four
Rayleigh \lgss, to allow investigation of extreme spot elongation, and
mitigation techniques (phase D).  Additionally, CANARY will host a
high order \scao upgrade, CHOUGH \citep{chough}.

\subsection{Phase A: Interfacing four NGS and a DM}
Initial designs for CANARY drew on experience gained by development of
the \eso \sparta \rtcs \citep{sparta}, including the ability to use a
modified \sparta-Light \wpu \citep{spartalight} for computation of
wavefront slopes in \fpga.  The \sfpdp protocol was used for real-time
communications.  The CANARY \rtcs is capable of operating on a
pixel-stream basis, rather than per-frame, i.e. processing commences
as soon as the first pixels arrive at the \rtcs, reducing the \ao
system latency.  To use this capability, which is key to good \ao
performance, it is necessary to have access to the pixel stream
produced by the camera, rather than (as with most commercial cameras)
receiving access on a per-frame basis.

\subsubsection{WFS cameras}
\begin{figure}
\includegraphics[width=0.65\linewidth]{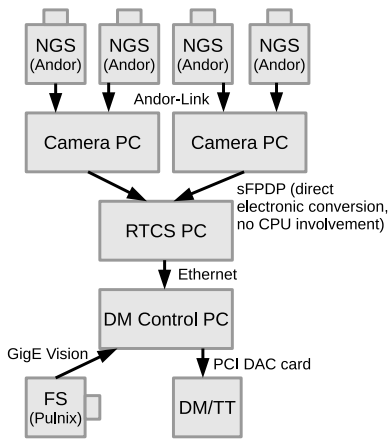}
\caption{Phase A of CANARY showing the camera and DM connectivity.  FS
represents the figure sensor, TT is the tip-tilt mirror, PC represents
a computer, NGS is a natural guide star wavefront sensor, RTCS is the
real-time control system and DM represents deformable mirror.}
\label{fig:phasea}
\end{figure}

The CANARY design (Fig.~\ref{fig:phasea}) selected four Andor
Technologies iXon 860 \emccd cameras, with a \pfpdp output for the pixel
stream, which was then converted to a fibre-based \sfpdp protocol with
pixels from pairs of cameras multiplexed together.  This data stream
(2 \sfpdp channels) was received using a commercial
\sfpdp PCI card in the \rtcs server.  The maximum frame
rate achievable with these cameras was 300~Hz.  Camera control
(cooling, frame rate, triggering, etc) was not via \sfpdp; instead,
the Andor Technologies \sdk and PCI control card were used.

\subsubsection{Deformable mirror}
The \dm and tip-tilt mirror were controlled using a 96 channel PCI
\dac card from United Electronic Industries, with custom electronics
for voltage conversion.  The \dm itself was reused from the ADONIS \ao
system \citep{adonis}, with 52 actuators ($8\times8$) and recently
characterised at LESIA \citep{2012SPIE.8447E..65K}.  The tip-tilt
mirror was made by Observatoire de Paris.  A schematic
diagram for phase A operation is given in Fig.~\ref{fig:phasea}

\subsubsection{Camera and DM control}
A camera interface library module for \darc was developed to receive
the \sfpdp pixel streams.  The \darc \rtcs pipeline then
de-multiplexed the pixel streams, calibrated the images, computed the
wavefront slopes and performed wavefront reconstruction and \dm
fitting (using a single control matrix).  Comprehensive lists of
computational algorithms used in CANARY (and DRAGON) are given by
\citet{basden9,basden11,dragonrt}.  The \dm command vector was then
output by \sfpdp to a \dmc server. 

\ignore{We note that although the CANARY \wpu was capable of calibrating
images and computing wavefront slopes, this feature was never used in
CANARY: the \cpu-based \rtcs had enough computational power to perform
these tasks, additional algorithms could easily be added, and system
complexity was therefore reduced.  }

\subsubsection{A figure sensor for open-loop DM control}
The \dm was operated in open loop, i.e.\ changes to the \dm surface
shape were not seen by the \wfss.  The \dmc included a figure sensor (a Shack-Hartmann
\wfs) was used to monitor the
shape of the \dm, and then apply offsets to the \dm command vectors so
that the actual shape matched the requested shape as closely as
possible.  The figure sensor was to operate at a significantly higher
frame rate than the \ngs \wfss, so that the \dm would move to its
correct shape over the course of one \ngs frame.

The figure sensor used a JAI Pulnix TM-6740GE camera, which had a GigE
Vision interface (directly connected to the \dmc), operating at up to
1~kHz frame rate for our region of interest.  A \darc interface
library was developed based on the camera manufacturer \sdk. 

\ignore{During initial verification and integration tests, it was determined
that the linearity of the \dm was excellent, and that the
figure sensor was not required.  It was therefore not used, except
to demonstrate that the technique worked.}

\subsubsection{Interface of the infrared science camera}
A Xenics Xeva-1.7-320 camera was selected as a science \psf
measurement camera, with a maximum frame rate of 60~Hz.  An interface
module for this \usb camera was created to allow operation with the
\rtcs, and it was operated with a separate instance of the \rtcs on a
dedicated science computer.  Integrating this camera with the \rtcs
gave us several key abilities: the tools for operation of the camera
and display and capture of information (locally and remotely) were
identical to the rest of the system.  Users therefore did not need to
learn additional interfaces.  This camera was also used to provide
tip-tilt closed loop \ao control during system calibration of the
\ncpas.

\subsubsection{On-sky operation}
For phase A of CANARY, we had two on-sky observing runs of 4 nights
each.  We were able to successfully demonstrate first \moao operation,
the capability of the Learn and Apply algorithm \citep{learnapply},
and first on-sky demonstration of \gpus for wavefront reconstruction.

\subsection{Phase B0: LGS commissioning}
Two pulsed 532~nm 20~Watt lasers were installed behind the \wht
secondary mirror, to provide a single Rayleigh \lgs using a
polarisation beam combiner.  Further details are given by
\citet{2011aoel.confP...3Mshort}.  During operation, these lasers are
pulsed at about 10~kHz, requiring a camera with a fast shutter to open
and close once each pulse has propagated to and returned from the
desired \lgs height.  Typically, this shutter will be open for
1~$\mu$s per pulse with 100~$\mu$s between pulses, ruling out any
practical mechanical shutter.  Therefore, a novel \ccd architecture
with an electronic (on-silicon) shutter (Lincoln Labs CCID18) was used
with a Scimeasure controller and a custom \sfpdp pixel stream
interface.  Unfortunately, this detector suffered electronic damage
shortly before the on-sky commissioning, and so a temporary solution
using Pockels cells and an Andor Technologies iXon camera was
implemented whilst a replacement detector was sourced.

During \lgs commissioning a PCO.Edge \scmos camera was used to provide
a wide field image of the laser (unshuttered), and was interfaced with
the \darc \rtcs so that common software tools could be used, including
pixel displays and camera controls, reducing duplication of effort.
Pixel stream acquisition was not enabled for this camera: since it was
not used in an \ao loop, this was not required.

The two lasers were polarisation combined to provide a single beam
with increased power.  A beam combining camera was installed to ensure
that alignment of the lasers was maintained.  This camera was an IDS
Imaging UI-2210SE VGA camera, which was also integrated with the \darc
\rtcs.  A separate instance of \darc was used for this camera,
i.e.\ it was not associated with the main \ao loop.  The ability to
automatically control the laser beam alignment was available because
of the \rtcs component.  A closed-source \sdk was used to create the
\darc interface module.

\subsubsection{On-sky operation}
Phase B0 of CANARY received a total of three on-sky commissioning runs
(a total of 9 nights) in 2011--2012.

\subsection{Phase B1: Single LGS AO operation}

\begin{figure}
\includegraphics[width=0.8\linewidth]{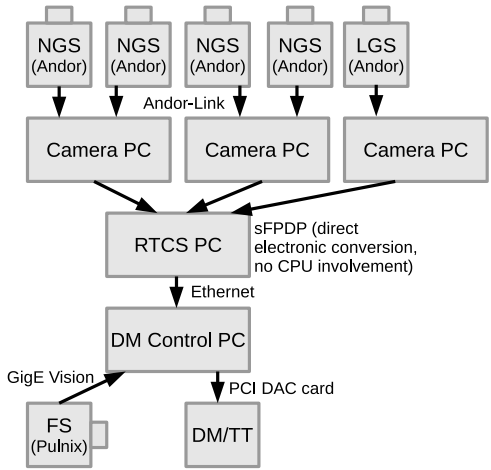}
\caption{Phase B1 of CANARY showing the camera and DM connectivity.
  Here, LGS represents the laser guide star wavefront sensor.}
\label{fig:phaseb1}
\end{figure}

After successful installation and testing of the lasers at the \wht,
CANARY proceeded with the operation of a single \lgs and 4 \ngss in
2012, as shown in Fig.~\ref{fig:phaseb1}.  At this phase of CANARY,
first successful on-sky demonstration of full \lqg \scao operation was
demonstrated \citep{lqgOptExpshort}.

A custom infrared science camera using a NICMOS
detector was introduced, replacing the previous science camera.  This
new detector had lower readout noise, and could operate with longer
exposure times.  An interface module for the \darc \rtcs was
developed, based on a \usb interface.

\subsubsection{On-sky operation}
Phase B1 of CANARY was operated with three on-sky observing runs (a
total of 12 nights) in 2012.

\subsection{Phase B2: Operation with 4 LGS}
\begin{figure}
\includegraphics[width=0.8\linewidth]{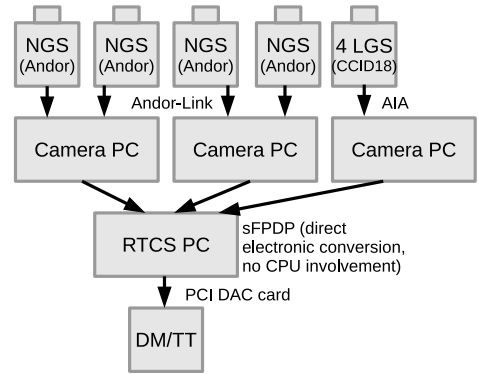}
\caption{Phase B2 of CANARY showing the camera and DM connectivity.}
\label{fig:phaseb2}
\end{figure}

Four \lgss were imaged onto four quadrants of the repaired
Scimeasure/CCID18 \wfs camera, each with $7\times7$ sub-apertures.  A
newer model of \ngs \wfs meant that maximum frame rate could now reach
450~Hz.  A schematic diagram is shown in Fig.~\ref{fig:phaseb2}.  At
this phase of CANARY, on-sky demonstration of tomographic wavefront
reconstruction using artificial neural networks was first demonstrated
\citep{Osborn01072014}, together with first demonstration of
tomographic \lqg control \citep{lqgao4elt}.

\subsubsection{On-sky operation}
Two on-sky observing runs of 6 nights each were used for this phase
 of CANARY (B2) in 2013, following an initial 3 nights for \lgs integration.

\subsection{Phase C1: LTAO operation}
\begin{figure}
\includegraphics[width=0.8\linewidth]{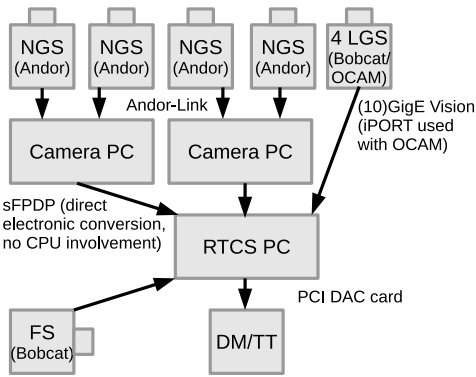}
\caption{Phase C1 of CANARY showing the camera and DM connectivity.}
\label{fig:phasec1}
\end{figure}

CANARY was
reconfigured into an \ltao system, by placing the \dm in closed loop
with the \wfss.  At this stage, the \lgs \wfs camera had again failed,
and a replacement was sought at short notice (1 month before on-sky
operation).  An Imperx Bobcat B0620 VGA camera was selected, using the
interline transfer region as an electronic shutter.  This camera had a
GigE Vision interface.  Readout noise restricted \lgs operation to about 12~km altitude (\lgs return
signal decreases rapidly with altitude).  Mid-way through phase C1
(during the second set of on-sky nights), this \lgs \wfs was replaced
by a First Light Imaging OCAM2S camera with developments specifically
for CANARY \citep{ocams}, and a schematic diagram is
shown in Fig.~\ref{fig:phasec1}.  The new camera allowed us to
increase \lgs altitude to about 20~km, though readout problems meant
that the \emccd gain mechanism was restricted.  

At this phase, computation of pseudo open-loop slopes became
necessary.  To aid this, a \dm figure sensor was installed, this
time using a Bobcat camera, operating as part of the main \rtcs loop.

\subsubsection{Integration of generic GigE vision cameras}
The necessity of pixel stream access meant that rather than using the
closed-source Imperx \sdk, we instead opted to use an open source GigE vision
library (Aravis) with modification to provide low
latency pixel stream access, i.e.\ the ability to access \udp packets
(and thus begin data processing) as soon as they arrive, rather than
waiting for the full frame.

The GigE Vision protocol is based on \udp packets, which are
inherently unreliable: the Linux Kernel can drop packets if processing
load gets too high.  Therefore, after investigation, it was determined
that to operate these cameras reliably, a hard real-time kernel was
necessary, and that the compute thread responsible for reading camera
pixels must be restricted to run on CPU cores directly attached to the
network card (by setting the thread affinity), with an elevated
priority.  After these steps are taken, \udp packet loss became
negligible, with less than one packet loss per hour.  The GigE Vision
interface module for the \darc system is suitable for operation with
any GigE Vision camera.

\subsubsection{Integration of the OCAM2S LGS WFS}
Due to the urgent need to replace the Bobcat camera because of high
readout noise, and to the very limited time available, a pragmatic
approach was taken to integrate the OCAM2S camera.  This camera has a
Camera Link interface, and was (for the frame grabbers that we had
available) restricted to full frame operation, i.e. pixel stream
access was not possible.  We therefore used a Camera Link to 10GigE
Vision converter (an iPort CL-Ten from Pleora), and so could use our
existing GigE Vision interface module for \darc.  The additional
latency introduced by the iPort was negligible, at the micro-second
level, and the maximum CANARY frame rate remained restricted by the
\ngs \wfs.

\subsubsection{On-sky operation}
Two on-sky observing runs (separated by about 3 months) of 6 nights
each were used for this phase of CANARY (C1) in 2014.

\subsection{Phase C2: Operation at increased WFS order and a woofer-tweeter configuration} 
\begin{figure}
\includegraphics[width=0.8\linewidth]{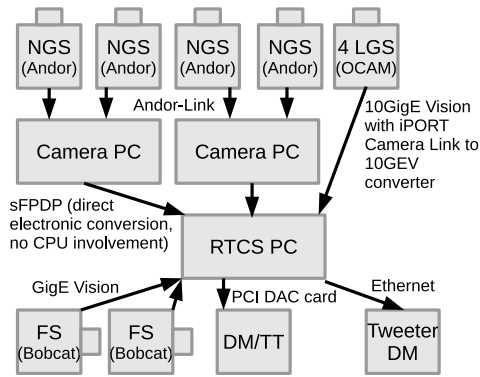}
\caption{Phase C2 of CANARY showing the camera and DM connectivity.}
\label{fig:phasec2}
\end{figure}

In 2015, CANARY was upgraded to provide woofer-tweeter control
\citep{woofertweeter} and a higher order \lgs \wfss and \scao \wfs
(the Truth sensor).  To achieve this, an additional \dm was added to
the system (Fig.~\ref{fig:phasec2}) in open-loop, (i.e.\ the \wfss
were insensitive to changes of the \dm surface), along with a
corresponding figure sensor (a Bobcat).  This tweeter \dm was an ALPAO
DM241, with $17\times17$ actuators.  The \lgs \wfs order was increased
from $7\times7$ sub-apertures to $14\times14$, again using the OCAM2S
camera, this time with full functionality and sub-electron effective
readout noise, allowing \lgs height to be increased to about 30~km.

It should be noted that this configuration is deliberately similar to
that proposed for the MOSAIC instrument \citep{2013arXiv1303.0029Eshort} on
the \eelt: A closed loop \dm (telescope M4) to provide \glao
correction, and open-loop \moao \dms.  This demonstration was the
original goal of CANARY, as proposed in 2007.

\subsubsection{Integration of the DM241}
The new \dm had an Ethernet interface and a closed-source binary \sdk.
Initial tests showed that there was some latency introduced by this
interface, equal to about 800~$\mu$s between \dm demands being
provided to the \sdk and the mirror surface settling.  For CANARY, the
additional latency was not critical due to pseudo-open-loop operation,
and we did not have time to develop an alternative solution.

To integrate the \dm Ethernet interface with CANARY, a new interface
module for \darc was developed, to allow operation of the woofer and
tweeter together.  The commissioning period of the upgraded system
prior to on-sky operation was limited to about 1 month.  However, due
to the flexibility afforded by the \darc interface module system,
integration of the new \dm was straightforward and no problems arose.

\subsubsection{On-sky operation}
CANARY phase C2 operated with two on-sky observing runs of 6 nights
each in 2015.

\subsection{Phase D: Extreme sodium LGS elongation}
Phase D of CANARY is planned for the second half of 2016 \citep{canaryD}, and will
replace the 4 Rayleigh \lgs with a single sodium \lgs, launched about
40~m from the telescope optical axis.  During this phase, extreme spot
elongation effects will be studied, with techniques developed to
mitigate spot truncation and loss of sensitivity along the elongation
axis.  Designs for the \eelt include sub-apertures that are about 40~m
from the laser launch axis.  Therefore, during CANARY phase D, we are
treating the \wht as a sub-pupil of the \eelt where \lgs elongation is
greatest.  Two observing runs are planned, of 6 nights each.

During phase D, we will use the existing CANARY \wfs cameras, each
with $7\times7$ sub-apertures.  The \lgs sub-apertures will have a
field of view of 20~arcseconds, to minimise spot truncation.  We will
also operate a \lgs profiling camera on the nearby \INT, which will
use a high resolution \scmos camera to image the \lgs profile,
allowing us to obtain high resolution sodium layer profile images both
to derive correlation and matched filter references, and for further
study of sodium layer variation.  The existing \scmos interface to the
\darc \rtcs will be used with modifications to allow dynamic resizing
of the detector region of interest.  A very simple 3-mode active
optics system will be implemented allowing control of tip, tilt and
rotation (to keep the \lgs elongation axis aligned with the detector
pixels).

\subsection{CANARY hosted high order upgrade}
The installation and test of a high order \ao instrument, CHOUGH
\citep{chough}, is also planned for 2016.  This will see CANARY
operating in \scao mode with $31\times31$ sub-apertures at a 1~kHz
frame-rate.  The camera to be used at this stage is a N\"{u}v\"{u} HNu
$128\times128$ \emccd camera with a GigE Vision interface.  The
existing woofer-tweeter \dms will be used, along with a Boston
Micromachines Kilo \dm, described in \S\ref{sect:boston}.

\subsection{The DRAGON wide-field AO bench}
DRAGON aims to replicate CANARY concepts, to provide a single channel
\moao system with a woofer-tweeter \dm configuration, 4 \ngss and 4
\lgss each with $30\times30$ sub-apertures.  The \wfss are all GigE
Vision standard, with the \ngs \wfss using the same model of Imperx Bobcat
cameras as used by CANARY, and the \lgs \wfss using an Emergent Vision
Technologies HS2000 10GigE Vision camera.  The \darc GigE Vision
interface library is used with these cameras, and thus no new
developments were required.

As a real-time research system, DRAGON enables verification of
\darc developments using hardware acceleration including \gpus, and
also many-core architectures such as the Xeon Phi \citep{phibarr}, or POWER8
processors \citep{basden20}.  Currently, the \darc \rtcs has the
capability to use \gpu acceleration, either just for wavefront
reconstruction (first demonstrated on-sky in 2010 during CANARY phase
A), or for the whole \ao pipeline.  Such hardware acceleration
capabilities will provide the ability to service increased
computational demands from future algorithm development.

\subsubsection{DM integration}
\label{sect:boston}
The woofer \dm is a Xinetics 97 actuator \dm controlled using a 96
channel PCI DAC card (United Electronic Industries, as used by CANARY)
with the central actuator slaved to neighbours (since it is behind the
central obscuration).  The tweeter \dm is a 1020 actuator Boston
Micromachines Kilo \dm ($32\times32$ actuators) controlled using a
PCIe fibre optic interface card provided by the \dm manufacturer.  The
tip-tilt mirror is controlled using a 16 channel PCI DAC card (United
Electronic Industries).  A \darc interface library to control these
\dms was required and developed.

\subsection{Summary of WFS cameras used in CANARY and DRAGON}
Table~\ref{tab:summary} provides a summary of key information about
the \wfs cameras that we have interfaced with \darc for operation with
the CANARY and DRAGON wide-field \ao systems.
\begin{table*}
\begin{tabular}{p{2cm}p{2cm}p{1.5cm}p{5cm}p{2.5cm}p{1.5cm}}\hline
Model & Type & Phase & Information & Interface & Full source code\\ \hline 

Andor iXon 860 & \ngs, \lgs commissioning & A, B, C, D & \emccd.  PCI compatibility problems for \sfpdp daughter board
used to send pixel stream & Control via manufacturer PCI, pixels via
\sfpdp & Yes\\ \cline{0-0}

Pulnix TM-6740GE & figure sensor & A & \sdk unusable with modern Linux
systems & GigE Vision & No\\ \cline{0-0}

PCO.Edge 5.5 & \lgs commissioning & B0 & \scmos & Camera Link & No\\ \cline{0-0}

UI-2210SE & \lgs beam combination & B, C & & \usb-2 & No\\ \cline{0-0}

Xeva-1.7-320 & IR science & A & & \usb-2 & No\\ \cline{0-0}

Camicaz & IR science & B, C, D & Custom built & \usb-2 & Yes\\ \cline{0-0}

Scimeasure/ CCID18 & \lgs & B2 & Electronic silicon shutter & AIA to
sFPDP & No\\ \cline{0-0}

Bobcat B0620GE & \lgs, figure sensor, \ngs & C, D, DRAGON & C1 as
\lgs, \ngs for DRAGON & GigE Vision & Yes\\ \cline{0-0}

OCAM2S & \lgs & C1, C2 & Used with iPORT converter & 10GigE Vision &
Yes\\ \cline{0-0}

HS-2000 & \lgs & DRAGON & & 10GigE Vision & Yes\\  \cline{0-0}

HNu 128 & High order \scao & CHOUGH & & GigE Vision & Yes\\ 
\hline

\end{tabular}
\caption{A table summarising key WFS parameters}
\label{tab:summary}
\end{table*}

\section{Challenges for WFS and DM integration}
The integration of such a diverse set of cameras and \dms using a
plethora of different interfaces meant that there were inevitable
challenges related to obtaining reliable operation, which we now discuss.

\subsection{Missing camera data and dropped frames}
For an \ao system with multiple camera inputs, it should be assumed
that at least occasionally, \wfs data will fail to arrive at the
\rtcs: even in the case of perfect hardware, random events such as
cosmic ray events could interfere.  During CANARY phase A
commissioning, we discovered that the \ngs \wfss would regularly
deliver partial frames, or insert an extra pixel into a frame.
Initially, this occurred about once every 4 minutes.  However, in
later phases of CANARY, the frequency increased, and became less
regular.  Our understanding is that this is due to the implementation
of the non-standard \pfpdp interface, and is beyond our control.
Therefore, a software fix, within the \rtcs, was required.

When using \udp-based cameras (i.e.\ GigE Vision), packet loss is
inevitable at some point, and so must be planned for.  

\subsubsection{Software handling of incorrect camera data}
Correct detection of incorrect camera data is key to \ao performance.
There are several cases that we have experienced, and thus consider
within the \darc \wfs interface modules.  

\begin{enumerate}
\item Dropped pixels: the data frame will be shorter than expected.
  This condition will be detected when the next start-of-frame signal
  arrives, after pipeline computation from all other \wfss has completed.
\item Inserted pixels (e.g.\ a single pixel being duplicated and thus
  received twice): the end-of-frame signal will be incorrect.
  This condition will be detected at the end of the frame, probably
  after pipeline computation has completed, i.e.\ wavefront
  reconstruction will have been performed based on corrupt data.
\item Complete frame missing: the \darc interface will not receive any
  data for this frame.  This condition will be detected when
  \ao pipeline computation for all other \wfs interfaces has completed
  and not yet started for the dropped frame.
\item Dropped Ethernet packets: \udp packets from GigE Vision cameras
  fail to arrive.  This condition is detected using the packet counter
  embedded within packets, and will be detected as soon as the next
  packet arrives, while pipeline computation is ongoing.  Within our
  interface, we do not make allowance for packet reordering since we
  have not discovered this occurring.  Therefore, an out-of-order
  packet is taken to signify a missing packet.
\end{enumerate}

\begin{figure}
\includegraphics[width=\linewidth]{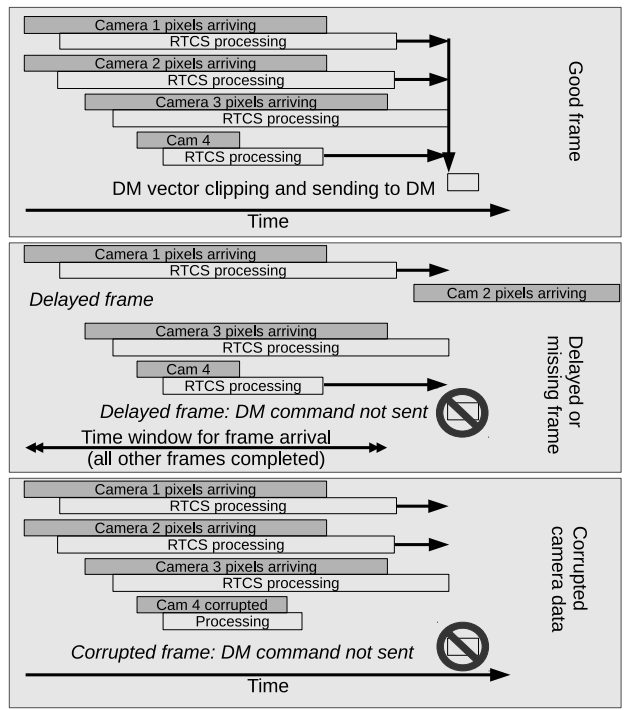}
\caption{A figure showing three cases for pixel arrival.  The top
  portion shows a good frame with all pixels arriving as expected.
  The middle case shows the action taken if one camera is delayed to
  beyond all others being finished (or fails to arrive).  The third
  case shows arrival of a corrupted frame.}
\label{fig:missing}
\end{figure}

These error conditions are detected and handled by \darc.  Upon
detection, a flag is set to specify that the \dm command vector should
be frozen for that frame, and that the integrator should not be
accumulated.  Fig.~\ref{fig:missing} demonstrates a time-line for
three cases.  During a good frame, the pixels from different camera
types arrive and are processed.  Note, different arrival times are due
to the differences in camera readout times (different camera models).
In the second case the arrival of a camera frame is delayed until
after all other cameras have finished readout, this is detected and
the pipeline aborted, i.e.\ \dm commands are not sent.  In the third
case a corrupted frame is detected in one camera, processing is aborted
and \dm commands are not sent.  

\begin{figure}
\includegraphics[width=\linewidth]{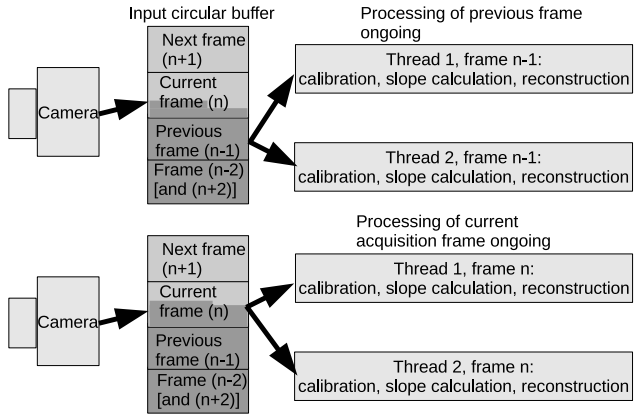}
\caption{A figure showing how pixel data is handled in DARC.  Camera
  pixels are directed to a small circular buffer which is then read by
the sub-aperture processing threads.  In the upper half, these threads
are processing a completed frame, and in the lower half, the threads
have moved on to begin processing the next, currently incomplete, frame.}
\label{fig:frames}
\end{figure}

Internally in \darc, one CPU thread is dedicated to each camera to
transfer pixels into a circular buffer for further processing by the CPU
threads responsible for \rtcs pipeline computation, which we term
sub-aperture processing threads.  Processing of these pixels proceeds
as soon as enough pixels to fill a given sub-aperture have arrived,
and a single thread then calibrates the sub-aperture, computes the
wavefront slope and performs a partial wavefront reconstruction.  We
term this a horizontal processing strategy, and further information is
given by \citet{basden9}.  The pixel arrival buffers are typically
quadruple buffered, i.e.\ a circular buffer with space for four camera
frames, though this depends on the \darc camera module being used.
Fig.~\ref{fig:frames} provides a schematic diagram of this approach.
Quadruple buffering will ensure that with a real-time kernel no data
is overwritten before it has been processed, and that with a
non-real-time kernel, this is also very unlikely.

Due to the use of this horizontal processing strategy, the detection
and mitigation of corrupted image frames is non-trivial: by the time
an error condition is detected, most of the computation for this and
other \wfss will already have been performed, including wavefront
reconstruction, update (decay) of the integrator, and update of
sub-aperture tracking algorithms (adaptive windowing).  It is
therefore necessary to reset these parameters to the previous state.
It is absolutely essential to ensure that \dm command vectors are
{\it never} sent to the \dm based on a corrupted camera frame.
Temporal forecasting from a previous frame is not a necessary strategy
for CANARY-scale systems: simply freezing the \dm state for a frame
does not significantly degrade performance (\S\ref{sect:corrupt}).

Maintaining synchronisation between the different \wfss (i.e. so that
one camera does not lag others by an integer number of frames) is
achieved by only using the most recent whole or partial frame after a
previous frame has finished.  As previously mentioned, one CPU thread
per camera is used to transfer pixels into a circular buffer.  When
starting a new frame, the sub-aperture processing threads first check
whether there is currently an active frame being read from the camera
(i.e.\ whether the first pixels have already arrived).  In this case,
this frame then proceeds to be processed.  If this is not the case
(i.e.\ there is not currently a frame arriving) then the most recently
acquired frame will be processed if it has not already been processed.
In this way, if a \wfs does begin to lag (if, for example, another
\wfs has missed a frame), a frame will be dropped, and differential
latency removed.  Although this approach might seem obvious, it is
worth a mention here as an issue that requires thought, i.e.\ is
non-trivial.  We note that this approach enables reliable operation on
non-real-time operating systems, i.e.\ scheduling delays do not allow
frames to stack up for processing.

\subsubsection{Impact of corrupt image frames}
\label{sect:corrupt}
\begin{figure}
\includegraphics[width=\linewidth]{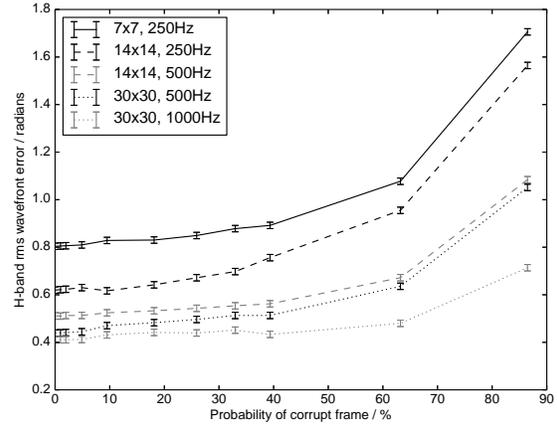}
\caption{A figure showing AO performance (rms residual wavefront
  error) as a function of probability of a corrupt WFS frame (for any
  given frame).  For
  CANARY, this probability is well below 1\%.}
\label{fig:fifo}
\end{figure}

The effect of missing or corrupt image frames on \ao performance for
CANARY is relatively low.  Being a moderate order (pseudo) open-loop
\ao system, \ao bandwidth error forms a relatively small part of the
overall error budget, and so the occasional additional single frame of
latency has little impact.  Fig.~\ref{fig:fifo} shows Monte-Carlo
simulation results (H-band rms wavefront error) for a
CANARY-like system (using only the on-axis Truth sensor) as a function of
probability of missing \wfs frames.  These results are for a \scao
system on a 4.2~m telescope and consider a $7\times7$, $14\times14$
and $30\times30$ order \ao system (Fried geometry), relevant for the different phases
of CANARY, and for CHOUGH and DRAGON, at different frame rates.  It
can be seen that when the probability of a missing frame is low (1\%),
the impact on performance is negligible.  For the CANARY \wfss, the
probability of a corrupt frame was found to be below the 0.1\% level,
and therefore we are confident that our occasional corrupt frame has
not reduced \ao performance with any significance: other errors
dominate.  Nevertheless, we recommend that \wfs cameras should be well
tested for these transient errors before acceptance.  

These Monte-Carlo simulations use parameters that are used during
CANARY design studies, including a 3-layer atmosphere with a Fried's
parameter of 12~cm and an outer scale of 30~m.  Within the simulation,
a corrupt image frame would be simulated with a specified probability
(Poisson distributed), which would then result in the \dm being frozen
for that frame.

Our technique for freezing system state within \darc (e.g.\ resetting
integrators to their previous values) upon detection of a corrupt
frame also applies to pseudo-open-loop slope calculation.

\ignore{
We note however that dropped frames (missing or corrupted) would lead
to significant performance degradation for \xao systems such as SPHERE
\citep{sphere}, which operate at a much higher frame rate, and which
have to consider performance criteria other than Strehl ratio.  \xao
\wfs cameras therefore have to be very well tested for these transient
errors before acceptance.
 }

\subsection{Camera trigger synchronisation}
Cameras in both CANARY and DRAGON are externally triggered using a
common frequency trigger signal, though with a selectable delay for
each camera.  For most phases of CANARY, this delay was set so that
the last pixel of each frame from each camera would arrive at the
\rtcs at the same time.  This approach minimised latency, allowing the
\dm shape to be set as soon as possible relative to the camera
exposures.  However, this approach means that different cameras are
exposed for different periods of time, leading to complications for
\polc operation during phase C.  At this point, we therefore triggered
cameras to have the same mid-point exposure time (i.e. the middle of
the exposure coincided for all cameras).  

Camera synchronisation can also be complicated by the inlcusion of
integrated electronic shutters that are used with the Rayleigh \lgs
\wfs detectors. Whilst this is by no means a standard technology used
within astronomy, the development of pulsed sodium lasers may mean
that it becomes a standard requirement. Dependent on the
implementation of the shuttering within the camera, the laser pulses
must also be synchronised to the camera readouts, requiring a
centralised timing system capable of nanosecond jitter feeding signals
to several distributed locations across the telescope.

\subsection{Camera driver issues}
Over the operational period of CANARY, we have had a large number of
different cameras interfaced to the CANARY \rtcs system.  Most of
these cameras have relied on closed-source software drivers, and as a
result we have experienced incompatibilities between required Linux
kernel specifications and software stacks, particularly for older
cameras which often do not see the related software updated for newer
Linux kernels.  

\begin{figure}
\includegraphics[width=\linewidth]{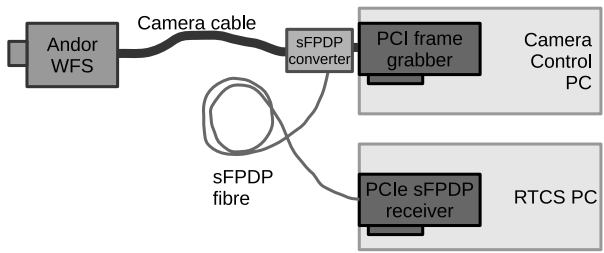}
\caption{A figure showing hardware used to access the pixel stream
  from Andor iXon cameras.}
\label{fig:dongle}
\end{figure}

For operation at phase A, we obtained (under a non-disclosure
agreement) source code for the \sfpdp receiver card used for capture
of \ngs pixels.  This was then essential at phase C when we upgraded
the \rtcs server, to allow the \sfpdp interface to continue to work
with a newer Linux kernel.  \emccd camera control was performed using
the standard camera interface card from Andor Technologies, which has
good driver support.  Unfortunately, our extension to enable a \sfpdp
pixel stream relies on a PCI card that we have only managed to operate
with one specific motherboard type, and of which our spare supplies
are running low.  We have therefore developed a new method for
producing the \sfpdp stream, using a \fpga based board which attaches
to the standard Andor Technologies camera output, acting as a pass-through device
for the standard image data, and also providing a \sfpdp (or Ethernet)
pixel stream, as shown in Fig.~\ref{fig:dongle}.  This system is
likely to be used from 2016 onward.

The Pulnix camera used a binary \sdk last updated in 2009, to which
source code access was not available.  Fortunately, this camera was
only required for phase A, and so future compatibility has not been an
issue.  Should we require use of this camera in the future, the newer
generic GigE Vision \darc interface (for which full source code is
available) would be used, also having the advantage of pixel stream access.

The \scmos camera has a Camera Link interface, and requires closed
source drivers for the frame grabber, as well as for the camera \sdk
(from different manufacturers).  As of 2015, these drivers have
remained in active development, and it has been possible to continue
to operate the camera with up to date Linux kernels.  However,
obtaining these drivers can be difficult.

The Scimeasure controller for the CCID18 detector uses an AIA frame
grabber card (necessary for camera control, even though we use \sfpdp
for receiving the pixel stream), requiring a PCI interface (which are
becoming less common).  Drivers are still available, though in binary
format from the frame grabber manufacturer.  A waveform compiler is
also necessary, and exists as a Windows executable (which we use on
Linux under Wine).  

As mentioned previously, our discovery of the open-source Aravis
library for GigE Vision cameras, and our modification of it meant that
any GigE Vision camera now has an interface to \darc which relies only
on the presence of a network interface, rather than a commercial frame
grabber.  In the case of the OCAM2S camera, a close collaboration with
the camera manufacturer, and an investigation of \udp packets was
necessary to develop a functional solution including full camera
control.

\subsection{Cabling of cameras}
The stability of an \ao bench is paramount and it is
necessary to keep electronic and computer racks some distance from the
optics to avoid heating, air-flow and vibration effects.  For CANARY, where
possible, computers and electronics are located in an electronics
room, adjacent to the optical bench area.

The available cable length for our Andor Technologies cameras was limited such that
we had the controller PCs mounted on an above-bench frame, with the
\sfpdp link extending from these PCs to the \rtcs (with fibre length
being essentially unlimited for our purposes).

For Ethernet based cameras, cable length is not an issue, as Ethernet
cable lengths are ample for our requirements.  Camera Link cables of
up to 10~m lengths are also available, again allowing a direct
connection between camera and the \rtcs in a separate electronics
room.  We have found that some Ethernet cameras get hot during
operation, which means thermal control should also be considered.

\ignore{
The Scimeasure controller cable was also just sufficient to allow the
camera controller to be located in the electronics room.  However, had
we required the camera at a different location on the optical table,
this would not have been possible.  
}

\subsection{DM driver issues}

The low order CANARY and DRAGON \dm and tip-tilt mirrors are
controlled using custom electronics driven from a commercial PCI \dac
card.  We have access to the source code for the drivers of this card,
which has allowed us to make modifications for newer Linux kernels.

The high order CANARY \dm is operated using Ethernet, though this
solution still required use of a closed source binary (at the user
level, i.e.\ without any kernel drivers necessary).

The Boston Kilo \dm uses a PCIe fibre card, for which closed-source
drivers are available for Linux.  Fortunately, the Kilo \dm drive electronics are modular, allowing a
different \dm interface to be used in the future, should the need arise.

\subsection{Lessons learnt and key points for consideration}
Our extensive experience with \wfss and \dms for wide-field \ao
systems has provided us with several key considerations to be taken
into account during the design of future \ao systems.  Many of the
problems that we encountered were specific to CANARY, however the
lessons that we learned are highly relevant for future \ao systems.
Closed source drivers and binary \sdks are problematic because of
potential future incompatibilities with newer Linux kernels due to
changes in the application binary interface specification, and should
be avoided where possible.  Systems using commonly available hardware
interfaces such as Ethernet should be favoured, and pixel stream
access will significantly reduce \ao latency.  Maximum cable lengths
should also be given consideration.  As astronomical \ao technology
becomes more mainstream, emphasis on commodity hardware and open
source software becomes increasingly important.

Synchronisation of \wfss and correct handling of corrupted image
frames is non-trivial due to the pipeline nature of \ao processing,
and should be considered at the design phase of \ao system
development.  The impact of corrupted frames on all aspects of the
system (telescope offloads, telemetry data storage, etc) should be
considered.

A single \rtcs system with which to operate all cameras has also been
beneficial (including cameras that are not \wfss), allowing a single
interface to be used, significantly reducing the learning curve for system
developers.  This also reduces the effort required to develop camera
control tools, graphical interfaces, etc, and simplifies project development.

\section{Conclusions}
The CANARY \ao demonstrator instrument has been operated on-sky over a
six year period, with many different instrument development phases,
aimed at testing and demonstrating new \ao concepts and technologies.
During this period we have acquired significant expertise related to
integration of \wfs cameras and \dms with the \ao \rtcs, \darc.  Here, we
have described the different phases of CANARY operation, providing
details of the \wfs and \dm interfaces required at each phase, and how
these have been integrated with the system.  An overview of the DRAGON
\ao bench has also been given, along with the approach taken for
integration of \wfss and \dms with the \rtcs.  We have discussed the
problems that were met and overcome, and have provided recommendations
for future \ao systems.  In summary, for long-life expectancy \ao
systems, we recommend the use of Ethernet based cameras and \dms where
possible to extend operational instrument lifetime, to enable
continued compatibility during future system updates, and to remove
the requirement for product specific frame grabbers or other hardware.
Open-source software, or as a minimum, access to source code for all
kernel module driver
interfaces greatly increases the future maintainability of these
systems, allowing continued developments, updates and repairs to be
made.

\section*{Acknowledgements}
This work is funded by the UK Science and Technology Facilities
Council, grant ST/K003569/1, and a consolidated grant ST/L00075X/1.  D.\ Guzman
appreciates support from FONDECYT grant 1150369.  Supported in France
by Agence Nationale de la Recherche (ANR) program
06-BLAN-0191, CNRS/INSU, Obs. de Paris and Univ. Paris Diderot ;
supported by European Commission FP7: E-ELT Prep. Infrastruct. 2007-1 Grant
211257, OPTICON program Infrastructures 2008-1 Grant 226604 (JRA1) and
2012-1 Grant 312430 (WP1)

\bibliographystyle{mn2e}

\bibliography{mybib}

\begin{thebibliography}{}

\bibitem[\protect\citeauthoryear{{Babcock}}{{Babcock}}{1953}]{adaptiveoptics}
{Babcock} H.~W.,  1953, \pasp, 65, 229

\bibitem[\protect\citeauthoryear{{Barr}, {Basden}, {Dipper} \&
  {Schwartz}}{{Barr} et~al.}{2015}]{phibarr}
{Barr} D.,  {Basden} A.~G.,  {Dipper} N.,    {Schwartz} N.,  2015, \mnras, 453,
  3222

\bibitem[\protect\citeauthoryear{{Basden}}{{Basden}}{2015}]{basden20}
{Basden} A.~G.,  2015, \mnras, 432, 1694

\bibitem[\protect\citeauthoryear{{Basden}, {Bharmal}, {Bitenc}, {Dipper},
  {Morris}, {Myers}, {Reeves} \& {Younger}}{{Basden} et~al.}{2014}]{dragonrt}
{Basden} A.~G.,  {Bharmal} N.~A.,  {Bitenc} U.,  {Dipper} N.,  {Morris} T.,
  {Myers} R.,  {Reeves} A.,    {Younger} E.,  2014, in Society of Photo-Optical
  Instrumentation Engineers (SPIE) Conference Series Vol.~9148 of Society of
  Photo-Optical Instrumentation Engineers (SPIE) Conference Series, {Real-time
  control for the high order, wide field DRAGON AO test bench}.
p.~4

\bibitem[\protect\citeauthoryear{{Basden}, {Geng}, {Myers} \&
  {Younger}}{{Basden} et~al.}{2010}]{basden9}
{Basden} A.~G.,  {Geng} D.,  {Myers} R.,    {Younger} E.,  2010, Appl.\ Optics,
  49, 6354

\bibitem[\protect\citeauthoryear{{Basden} \& {Myers}}{{Basden} \&
  {Myers}}{2012}]{basden11}
{Basden} A.~G.,  {Myers} R.~M.,  2012, \mnras, 424, 1483

\bibitem[\protect\citeauthoryear{{Bharmal}, {Myers}, {Basden}, {H{\"o}lck} \&
  {Morris}}{{Bharmal} et~al.}{2014}]{chough}
{Bharmal} N.~A.,  {Myers} R.~M.,  {Basden} A.~G.,  {H{\"o}lck} D.,    {Morris}
  T.~J.,  2014, in Society of Photo-Optical Instrumentation Engineers (SPIE)
  Conference Series Vol.~9148 of Society of Photo-Optical Instrumentation
  Engineers (SPIE) Conference Series, {CHOUGH, the Canary Hosted-Upgrade for
  High-Order Adaptive Optics}.
p.~5

\bibitem[\protect\citeauthoryear{{Evans}, {Puech}, {Barbuy}, {Bastian},
  {Bonifacio}, {Caffau}, {Cuby}, {Dalton}, {Davies}, {Dunlop}, {Flores},
  {Hammer}, {Kaper}, {Lemasle} \& {Morris}}{{Evans}
  et~al.}{2013}]{2013arXiv1303.0029Eshort}
{Evans} C.,  {Puech} M.,  {Barbuy} B.,  {Bastian} N.,  {Bonifacio} P.,
  {Caffau} E.,  {Cuby} J.-G.,  {Dalton} G.,  {Davies} B.,  {Dunlop} J.,
  {Flores} H.,  {Hammer} F.,  {Kaper} L.,  {Lemasle} B.,    {Morris} S.,  2013,
  ArXiv e-prints

\bibitem[\protect\citeauthoryear{{Fedrigo}, {Donaldson}, {Soenke}, {Myers},
  {Goodsell}, {Geng}, {Saunter} \& {Dipper}}{{Fedrigo} et~al.}{2006}]{sparta}
{Fedrigo} E.,  {Donaldson} R.,  {Soenke} C.,  {Myers} R.,  {Goodsell} S.,
  {Geng} D.,  {Saunter} C.,    {Dipper} N.,  2006, in Advances in Adaptive
  Optics II. Edited by Ellerbroek, Brent L.; Bonaccini Calia, Domenico.
  Proceedings of the SPIE, Volume 6272, pp. 627210 (2006). Vol.~6272 of
  Presented at the Society of Photo-Optical Instrumentation Engineers (SPIE)
  Conference, {SPARTA: the ESO standard platform for adaptive optics real time
  applications}

\bibitem[\protect\citeauthoryear{{Gach}, {Feautrier}, {Buey}, {Rousset},
  {Gendron}, {Morris}, {Basden}, {Stadler}, {Myers}, {Vidal} \&
  {Chemla}}{{Gach} et~al.}{2016}]{ocams}
{Gach} J.-L.,  {Feautrier} P.,  {Buey} T.,  {Rousset} G.,  {Gendron} E.,
  {Morris} T.~J.,  {Basden} A.,  {Stadler} E.,  {Myers} E.,  {Vidal} F.,
  {Chemla} F.,  2016, in {Gavel} D.,  {Trancho} G.,  eds, Proceedings of the
  Fourth AO4ELT Conference {OCAM2S: an integral shutter ultrafast and low noise
  wavefront sensor camera for laser guide stars adaptive optics systems}.
p.~58

\bibitem[\protect\citeauthoryear{{Gendron}, {Vidal}, {Brangier}, {Morris},
  {Hubert}, {Basden}, {Rousset} \& {Myers}}{{Gendron}
  et~al.}{2011}]{canaryresultsshort}
{Gendron} E.,  {Vidal} F.,  {Brangier} M.,  {Morris} T.,  {Hubert} Z.,
  {Basden} A.~G.,  {Rousset} G.,    {Myers} R.,  2011, \aap, 529, L2

\bibitem[\protect\citeauthoryear{{Hampton}, {Bradley}, {Agathoklis} \&
  {Conan}}{{Hampton} et~al.}{2006}]{woofertweeter}
{Hampton} P.,  {Bradley} C.,  {Agathoklis} P.,    {Conan} R.,  2006, in The
  Advanced Maui Optical and Space Surveillance Technologies Conference {Control
  System Performance of a Woofer-Tweeter Adaptive Optics System}.
p.~59

\bibitem[\protect\citeauthoryear{{Jagourel} \& {Gaffard}}{{Jagourel} \&
  {Gaffard}}{1992}]{adonis}
{Jagourel} P.,  {Gaffard} J.-P.,  1992, in {Ealey} M.~A.,  ed., Active and
  Adaptive Optical Components Vol.~1543 of Society of Photo-Optical
  Instrumentation Engineers (SPIE) Conference Series, {Adaptive optics
  components in Laserdot}.
pp 76--87

\bibitem[\protect\citeauthoryear{Johns}{Johns}{2008}]{gmt}
Johns M.,  2008, in Extremely Large Telescopes: Which Wavelengths? Retirement
  Symposium for Arne Ardeberg Vol.~6986, {The Giant Magellan Telescope (GMT)}.
pp 698603--698603--12

\bibitem[\protect\citeauthoryear{{Kellerer}, {Vidal}, {Gendron}, {Hubert},
  {Perret} \& {Rousset}}{{Kellerer} et~al.}{2012}]{2012SPIE.8447E..65K}
{Kellerer} A.,  {Vidal} F.,  {Gendron} E.,  {Hubert} Z.,  {Perret} D.,
  {Rousset} G.,  2012, in Adaptive Optics Systems III Vol.~8447 of Society of
  Photo-Optical Instrumentation Engineers (SPIE) Conference Series, {Deformable
  mirrors for open-loop adaptive optics}.
p. 844765

\bibitem[\protect\citeauthoryear{{Morris}, {Hubert}, {Chemla}, {Todd},
  {Gendron}, {Huet}, {Younger} \& {Basden}}{{Morris}
  et~al.}{2011}]{2011aoel.confP...3Mshort}
{Morris} T.,  {Hubert} Z.,  {Chemla} F.,  {Todd} S.,  {Gendron} E.,  {Huet}
  J.-M.,  {Younger} E.,    {Basden} A.~G.,  2011, in Second International
  Conference on Adaptive Optics for Extremely Large Telescopes. Online at
  http://ao4elt2.lesia.obspm.fr, id.P3 {CANARY Phase B: the LGS upgrade to the
  CANARY tomographic MOAO pathfinder}.
p.~3P

\bibitem[\protect\citeauthoryear{{Myers}, {Hubert}, {Morris}, {Gendron},
  {Dipper}, {Kellerer}, {Goodsell}, {Rousset}, {Younger}, {Marteaud} \&
  {Basden}}{{Myers} et~al.}{2008}]{canaryshort}
{Myers} R.~M.,  {Hubert} Z.,  {Morris} T.~J.,  {Gendron} E.,  {Dipper} N.~A.,
  {Kellerer} A.,  {Goodsell} S.~J.,  {Rousset} G.,  {Younger} E.,  {Marteaud}
  M.,    {Basden} A.~G.,  2008, in Society of Photo-Optical Instrumentation
  Engineers (SPIE) Conference Series Vol.~7015 of Presented at the Society of
  Photo-Optical Instrumentation Engineers (SPIE) Conference, {CANARY: the
  on-sky NGS/LGS MOAO demonstrator for EAGLE}

\bibitem[\protect\citeauthoryear{{Nelson} \& {Sanders}}{{Nelson} \&
  {Sanders}}{2008}]{tmt}
{Nelson} J.,  {Sanders} G.~H.,  2008, in Society of Photo-Optical
  Instrumentation Engineers (SPIE) Conference Series Vol.~7012 of Society of
  Photo-Optical Instrumentation Engineers (SPIE) Conference Series, {The status
  of the Thirty Meter Telescope project}.
pp 70121A--70121A--18

\bibitem[\protect\citeauthoryear{Osborn, Guzman, de Cos~Juez, Basden, Morris,
  Gendron, Butterley, Myers, Guesalaga, Sanchez~Lasheras, Gomez~Victoria,
  Sánchez~Rodríguez, Gratadour \& Rousset}{Osborn
  et~al.}{2014}]{Osborn01072014}
Osborn J.,  Guzman D.,  de Cos~Juez F.~J.,  Basden A.~G.,  Morris T.~J.,
  Gendron E.,  Butterley T.,  Myers R.~M.,  Guesalaga A.,  Sanchez~Lasheras F.,
   Gomez~Victoria M.,  Sánchez~Rodríguez M.~L.,  Gratadour D.,    Rousset G.,
   2014, Monthly Notices of the Royal Astronomical Society, 441, 2508

\bibitem[\protect\citeauthoryear{{Reeves}, {Myers}, {Morris}, {Basden},
  {Bharmal}, {Rolt}, {Bramall}, {Dipper} \& {Younger}}{{Reeves}
  et~al.}{2012}]{dragon}
{Reeves} A.~P.,  {Myers} R.~M.,  {Morris} T.~J.,  {Basden} A.~G.,  {Bharmal}
  N.~A.,  {Rolt} S.,  {Bramall} D.~G.,  {Dipper} N.~A.,    {Younger} E.~J.,
  2012, in Society of Photo-Optical Instrumentation Engineers (SPIE) Conference
  Series Vol.~8447 of Society of Photo-Optical Instrumentation Engineers (SPIE)
  Conference Series, {DRAGON: a wide-field multipurpose real time adaptive
  optics test bench}

\bibitem[\protect\citeauthoryear{{Rigaut}, {Neichel}, {Boccas}, {d'Orgeville},
  {Arriagada}, {Fesquet}, {Diggs}, {Marchant}, {Gausach}, {Rambold}, {Luhrs},
  {Walker}, {Carrasco-Damele}, {Edwards}, {Pessev} \& {Galvez}}{{Rigaut}
  et~al.}{2012}]{2012SPIE.8447E..0IRshort}
{Rigaut} F.,  {Neichel} B.,  {Boccas} M.,  {d'Orgeville} C.,  {Arriagada} G.,
  {Fesquet} V.,  {Diggs} S.~J.,  {Marchant} C.,  {Gausach} G.,  {Rambold}
  W.~N.,  {Luhrs} J.,  {Walker} S.,  {Carrasco-Damele} E.~R.,  {Edwards} M.~L.,
   {Pessev} P.,    {Galvez} R.~L.,  2012, in Society of Photo-Optical
  Instrumentation Engineers (SPIE) Conference Series Vol.~8447 of Society of
  Photo-Optical Instrumentation Engineers (SPIE) Conference Series, {GeMS:
  first on-sky results}

\bibitem[\protect\citeauthoryear{{Rousset}, {Gratadour}, {Gendron}, {Buey},
  {Myers}, {Morris}, {Basden}, {Talbot}, {Bonaccini Calia}, {Marchetti} \&
  {Pfrommer}}{{Rousset} et~al.}{2014}]{canaryD}
{Rousset} G.,  {Gratadour} D.,  {Gendron} E.,  {Buey} T.,  {Myers} R.,
  {Morris} T.,  {Basden} A.~G.,  {Talbot} G.,  {Bonaccini Calia} D.,
  {Marchetti} E.,    {Pfrommer} T.,  2014, in Society of Photo-Optical
  Instrumentation Engineers (SPIE) Conference Series Vol.~9148 of Society of
  Photo-Optical Instrumentation Engineers (SPIE) Conference Series, {Proposal
  for a field experiment of elongated Na LGS wave-front sensing in the
  perspective of the E-ELT}.
p.~3

\bibitem[\protect\citeauthoryear{Sivo, Kulcsar, Conan, Raynaud, Gendron,
  Basden, Vidal \& Morris}{Sivo et~al.}{2014}]{lqgOptExpshort}
Sivo G.,  Kulcsar C.,  Conan J.,  Raynaud H.,  Gendron E.,  Basden A.~G.,
  Vidal F.,    Morris T.,  2014, Opt.\ Express, 22, 23565

\bibitem[\protect\citeauthoryear{{Sivo}, {Kulcsar}, {Conan}, {Raynaud},
  {Gendron}, {Basden}, {Vidal}, {Morris}, {Meimon}, {Petit}, {Gratadour},
  {Martin}, {Hubert}, {Rousset}, {Dipper}, {Talbot}, {Younger} \&
  {Myers}}{{Sivo} et~al.}{2013}]{lqgao4elt}
{Sivo} G.,  {Kulcsar} C.,  {Conan} J.-M.,  {Raynaud} H.-F.,  {Gendron} E.,
  {Basden} A.~G.,  {Vidal} F.,  {Morris} T.,  {Meimon} S.,  {Petit} C.,
  {Gratadour} D.,  {Martin} O.,  {Hubert} Z.,  {Rousset} G.,  {Dipper} N.,
  {Talbot} G.,  {Younger} E.,    {Myers} R.,  2013, in {Esposito} S.,  {Fini}
  L.,  eds, Proceedings of the Third AO4ELT Conference {First on-sky validation
  of full LQG control with vibration mitigation on the CANARY MOAO pathfinder}.
p.~127

\bibitem[\protect\citeauthoryear{{Spyromilio}, {Comer{\'o}n}, {D'Odorico},
  {Kissler-Patig} \& {Gilmozzi}}{{Spyromilio} et~al.}{2008}]{eelt}
{Spyromilio} J.,  {Comer{\'o}n} F.,  {D'Odorico} S.,  {Kissler-Patig} M.,
  {Gilmozzi} R.,  2008, The Messenger, 133, 2

\bibitem[\protect\citeauthoryear{{Su{\'a}rez Valles}, {Fedrigo}, {Donaldson},
  {Soenke}, {Zampieri}, {Bourtembourg} \& {Tischer}}{{Su{\'a}rez Valles}
  et~al.}{2012}]{spartalight}
{Su{\'a}rez Valles} M.,  {Fedrigo} E.,  {Donaldson} R.~H.,  {Soenke} C.,
  {Zampieri} S.,  {Bourtembourg} R.,    {Tischer} H.,  2012, in Society of
  Photo-Optical Instrumentation Engineers (SPIE) Conference Series Vol.~8447 of
  Society of Photo-Optical Instrumentation Engineers (SPIE) Conference Series,
  {SPARTA for the VLT: status and plans}.
p.~2

\bibitem[\protect\citeauthoryear{{Vidal}, {Gendron}, {Brangier}, {Sevin},
  {Rousset} \& {Hubert}}{{Vidal} et~al.}{2010}]{learnapply}
{Vidal} F.,  {Gendron} E.,  {Brangier} M.,  {Sevin} A.,  {Rousset} G.,
  {Hubert} Z.,  2010, in Adaptative Optics for Extremely Large Telescopes
  {Tomography reconstruction using the Learn and Apply algorithm}

\end{thebibliography}
\bsp

\end{document}